\documentclass[aps,prd,reprint,groupedaddress,showpacs,showkeys,longbibliography]{revtex4-1}

\usepackage{graphicx}

\begin{document}


\title[A non-dynamical approach for quantum gravity]{A non-dynamical approach for quantum gravity}


\author{Pierre A. Mandrin}

\email[]{pierre.mandrin@uzh.ch}
\affiliation{Department of Physics, University of Zurich, Winterthurerstrasse 190, 8057 
Z\"urich, CH}


\date{\today}

\begin{abstract}
By quantising the gravitational dynamics, space and time are usually forced to play fundamentally different roles. This raises the question whether physically relevent configurations could also exist which would not admit space-time-splitting. This has led to the investigation of an approach not based on quantum dynamical assumptions. The assumptions are mainly restricted to a constrained statistical concept of ordered partitions (NDA). For the time being, the continuum description is restricted in order to allow the application of the rules of differential geometry. It is verified that NDA yields equations of the same form as general relativity and quantum field theory for 3+1 dimensions and within the limits of experimental evidence. The derivations are shown in detail. First results are compared to the path integral approach to quantum gravity.
\end{abstract}


\maketitle


\section{Introduction}
\label{intro}

The existing quantum gravity theories are mostly confronted with the incompatibility between general relativity (GR) and quantum mechanics (QM). GR and QM treat space and time in fundamentally different ways which exclude each other. GR excludes that space and time can play different roles as opposed to QM. According to canonical quantisation, the space-time splitting structure of QM is transferred to quantum gravity via the Hamilton-Jacobi formulation. This splitting might cause a loss of ''unsplitted'' but physically meaningful states. Such a possible loss of states could be excluded either by introducing polymomenta \cite{Kanatchikov} or, in the present article, by considering finite and ordered partitions of quantum objects not based on dynamics, then evaluating the probability for statistical distributions to occur. In order to put the model into relation with gravitational space, a parameterisation will be introduced. It will be convenient to focus on the case for which differential geometry methods apply.

\paragraph*{}
The procedure is motivated by two hints. The first hint comes from black hole thermodynamics. The simplest example is the Schwarzschild black hole which obeys physical laws analogous to a many-particle system in thermal equilibrium. The area of the horizon corresponds to the Bekenstein entropy \cite{Bekenstein}, while the surface gravity corresponds to a temperature causing Hawking radiation \cite{Hawking}. In this sense, the gravitational space might have a dual system obeyed by statistics.
The second hint comes from the similarities between the variation principle and the second law of thermodynamics. While the variation principle extremises the action of e.g. the gravitation field, the second law extremises its dual, the entropy of the statistical system. This latter correspondence has been the main motivation for the present investigations.

\paragraph*{}
The detailed procedure is as follows: In view of a thermodynamic treatment, we consider a statistically large number $N \gg \sqrt{N}$ of distinct objects of the dual system (let us call them ''primary quanta''). In order to analyse their collective properties, we will introduce partitions of the set of quanta into ordered subsets as well as a parameterisation tool, and then apply the second law of thermodynamics. Let us call this the non-dynamical approach (NDA).
There are superficially similar features between NDA and causal sets (or related theories) with abstract elements and ordering structures, but causal sets require a function (or ''Hamiltonian'') in order to filter out the unphysical sets and thus requires quantum dynamics. In contrast, no such quantum dynamics is required by NDA. There have also been efforts to obtain quantum gravity from thermodynamics, e.g. \cite{Brown_York_1992, Brown_York_1993, Creighton_Mann}, but all attempts so far either assume quantum dynamics of some kind or else have no quantum model description at all. Although the path integral method \cite{Hamber} provides a manifestly covariant model, the adequate choice of the measure remains an open question. Unfortunately, the measure is  related to the space of physically meaningful states.

\paragraph*{}
Although a similar non-dynamical procedure with a few more detours has been proposed before \cite{Mandrin1, Mandrin2, Mandrin2a,
Mandrin3a, Mandrin4}, the derivation differs and none of these former articles is required in order to understand the present article. This article is self-contained and contains new developments for the test of compatibility with GR and QFT. The article is written with emphasis on the conceptual ideas, so that the reader can follow every step safely. In the case of 3-dimensional parameterisation, it is shown that the dual of a generalisation of GR and the dual of QFT are obtained. Furthermore, it is shortly sketched how quantum phenomena may be described and quantified. As for any quantum gravity model, the scientific program would be far to ambitious to be presented as a whole in one single article. Further developments shall be presented in further articles, with the following goals: recover the 3 dimensions of the dual of the gravitational space, recover a model for dual matter fields and obtain a general structure of the interactions (the dual standard model should be a special case), obtain general solutions of NDA.

\section{Partitioning, ordering and equilibrium}
\label{sec:para}

\begin{figure}[hbtp]
\begin{center}
\includegraphics{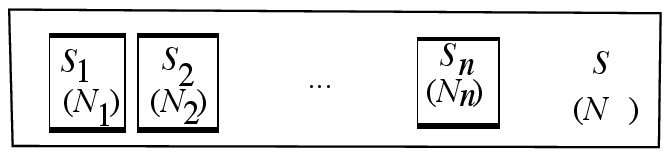}
\end{center}
\caption[Partitioning and ordering of S]{\label{fig:1}Partitioning and ordering of S}
\end{figure}

We begin with a statistically large number $N \gg \sqrt{N}$ of primary quanta. Suppose we wish to have insight on some part of these quanta. We then need to partition the total system $\mathcal{S}$ of $N$ quanta into some number $n$ of subsystems $\mathcal{S}_i$ containing $N_i$ quanta and labelled $i = 1 \ldots n$ (see figure \ref{fig:1}). Partition means $\mathcal{S}= \bigcup_i \mathcal{S}_i$ and the intersection of any pair $(\mathcal{S}_i$, $\mathcal{S}_{j\ne i})$ is empty. 

\paragraph*{}
The label $i$ allows us to define an ordering of the subsystems: Given any pair $(\mathcal{S}_i$, $\mathcal{S}_{j\ne i})$, either $\mathcal{S}_j$ is the covering of $\mathcal{S}_i$ ($\mathcal{S}_i <: \mathcal{S}_j$) or $\mathcal{S}_i$ is the covering of $\mathcal{S}_j$ ($\mathcal{S}_j <: \mathcal{S}_i$) or neither applies. The partitioning and the ordering can be performed in a fully arbitrary manner. Changing the ordering may be interpreted as a change of partitioning. The partitioning shall thus be denoted by $\mathcal{P} = (\{\mathcal{S}_i\}; <:)$. The numbers $N_i$ can change depending on the choice of $\mathcal{P}$ while the total number $N$ remains unchanged. Quantities which, like $N$, do not depend on $\mathcal{P}$ (while $\mathcal{S}$ is fixed) shall be called conserved quantities. Let us call a partition $\mathcal{P} = (\{\mathcal{S}_i\}; <:)$ macroscopic if every $\mathcal{S}_i$ has a large number of quanta $N_i \gg \sqrt{N_i}$. 

\begin{figure}[hbtp]
\begin{center}
\includegraphics{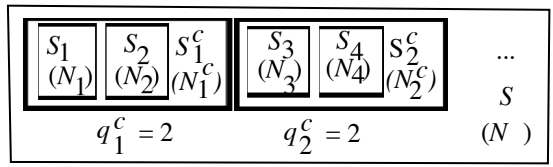}
\end{center}
\caption[Example of a coarse-grained partition]{\label{fig:2}Example of a coarse-grained partition}
\end{figure}

\paragraph*{}
The goal is to obtain information about the inner structure of $\mathcal{S}$, e.g. the numbers $N_i$ of the subsystems $\mathcal{S}_i$. As there are many possible partitions $\mathcal{P}$, we can evaluate the values $N_i$ in many different ways. Consider a sample of infinitely many evaluations of random partitions. Then, all possible arrays $(N_1,N_2, \ldots, N_n)$ occur. Each array  occurs with equal probability. Things change when we also consider coarse-grained partitions $\mathcal{P}^c$ with $n^c < n$  such that each $\mathcal{S}^c_j$ contains $N^c_j$ quanta and $q^c_j$ subsystems $\mathcal{S}_i$, $q^c_j=n/n^c \ge 2$ ($j=1\ldots n^c$), see figure \ref{fig:2}. Let us call the structure $\mathcal{D}^c=(\{N^c_j\}, \{q^c_j\})$ the distribution of $\mathcal{P}^c$ over the set of sub-partitions with $q^c_j$ subsystems per $\mathcal{S}^c_j$. For all arrays $\{N_i\}$ contained in $\mathcal{D}^c$, each $N^c_j$ equals the sum over the $N_i$ of its $q^c_j$ subsystems. The maximum number of arrays is obtained for a uniform distribution $\mathcal{D}^c$ with $N^c_j \approx N/n^c$. For given $N$, $n$, let me call $\mathcal{P}^c$ an equilibrium partition if $\mathcal{D}^c=(\{N^c_j\}, \{q^c_j=n/n^c\})$ is a uniform distribution (for arbitrary $n/n^c \ge 2$). The term ''equilibrium'' is chosen in analogy to statistical mechanics. The equilibrium partitions must be the preferred ones because they occur with the largest probability when we probe the $N^c_j$.

\paragraph*{}
In order to make the mathematics simple in what follows, let us split the statistical computations into two steps. In the first step, we evaluate the number of arrays of a distribution without any conservation condition on $N = \sum_i N_i$. In the second step, we impose the conservation condition $N = N_0 =$ constant as a constraint. Consider now the first step and partitions $\mathcal{P}^b$ with fixed number of subsystems $n^b$. In order that the number of possible partitions $\mathcal{P}^b$ be finite, we must set an upper limit $p-1$ on $N^b_i$, $N^b_i \le p-1$. The constant $p-1 \ge 1$ should be much smaller than any possible constraint $N_0 \gg \sqrt{N_0}$, i.e. $p-1$ is of order of 1 or slightly more than 1 and otherwise arbitrary. We must then hold $p$ fixed during the computations. Let us denote the subsystems $\mathcal{S}_i^b$ (with this property) as boxes. Therefore, the number of possible arrays for a partition $\mathcal{P}^b$ of $n^b$ boxes is $\Omega = p^{n^b}$, and the entropy is thus

\begin{equation}
\label{eq:entropy}
S = \ln{\Omega} = n^b \ \ln{p}.
\end{equation}

\noindent Let us rescale $N = \sum_i N^b_i \rightarrow E = \ln{p} \cdot N$. Let us define the quantity

\begin{equation}
\label{eq:temperature}
T = \frac{E}{S} = \frac{N}{n^b}.
\end{equation}

\begin{figure}[hbtp]
\begin{center}
\includegraphics{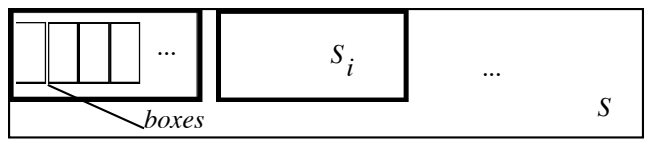}
\end{center}
\caption[Partition and boxes]{\label{fig:3}Partition and boxes}
\end{figure}

\noindent If we remove or add boxes, one also has $T = \delta E / \delta S$, so that we could interpret $T$ as a ''temperature''. We now can easily specify the pattern of $T$ for a macroscopic partition $\mathcal{P}$ (illustrated in figure \ref{fig:3}), replacing $T \rightarrow T_i, N \rightarrow N_i, n^b \rightarrow n_i$ in (\ref{eq:temperature}). If $\mathcal{P}$ is equilibrium, the values of $T_i$ are equal, $T_i = T$, and the distribution $(\{N_i\}, \{n_i\})$ is uniform.

\section{Macroscopic parameterisation of partitions}
\label{sec:param}

\begin{figure}[hbtp]
\begin{center}
\includegraphics{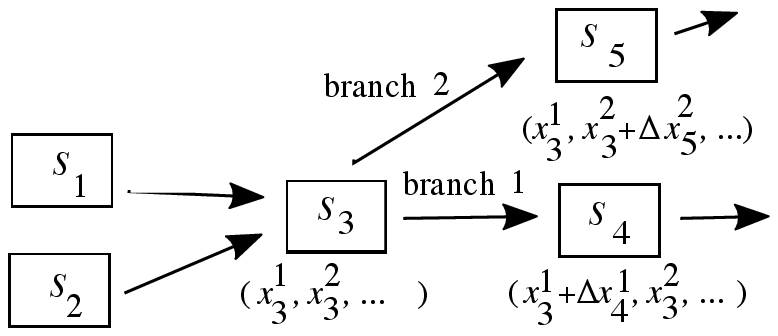}
\end{center}
\caption[Parameterisation of a very fine macroscopic partition]{\label{fig:4}Parameterisation of a very fine macroscopic partition}
\end{figure}

\paragraph*{}
Next, it is convenient to introduce a macroscopic parameterisation based on a macroscopic but very fine partition $\mathcal{P}$, as illustrated in figure \ref{fig:4}.
We attach arbitrarily chosen parameter values $x(\mathcal{S}_3)=(x_3^1, x_3^2, \ldots)$  to $\mathcal{S}_3$. For the covering $\mathcal{S}_4 :> \mathcal{S}_3$, we can add a positive amount $\Delta x_4^1$ to the first component $x_3^1$, $x(\mathcal{S}_4)=(x_3^1+\Delta x_4^1, x_3^2, \ldots)$. Further members of the branch 1 can be denoted in the same way by increasing the first component $x^1$. It is convenient to distinguish the branch 2 going through $\mathcal{S}_5$ from the branch 1 going through $\mathcal{S}_4$, by increasing the value of the independent component $x_3^2$: $x(\mathcal{S}_5)=(x_3^1, x_3^2+\Delta x_5^2, \ldots)$. Let us define $m_i$ as the number of different coverings of $\mathcal{S}_i$.

Every $\mathcal{S}_i$ plays a role similar to a point on a space-time manifold. Let us thus attach to every $\mathcal{S}_i$ a local vector space $V_i$ isomorphic to $R^{m_i}$ with notation in components $(x^1, \ldots x^{m_i})$. With the above prescription for coverings, $\mathcal{S}$ is a topological space. Moreover, arbitrary translations and rotations of $x^k$ do not affect the resulting quantum distributions. Finally, there is a freedom in the scaling of $x^k$, in analogy to the free calibration of space and time.

\paragraph*{}
It is easy to show that the values of the $m_i$ of a sufficiently fine equilibrium partition $\mathcal{P}$ with $n$ subsystems $\mathcal{S}_i$ do not depend on $i$. For an equilibrium partition, we not only have $N_i = N/n$ but also $m_i = D/n$, where $D$ is the total number of coverings contained in $\mathcal{P}$. This follows by the same arguments as for the values $N_i = N/n$. Therefore, $m_i = d =$ constant. We call $d$ the dimension. Also, starting from a subsystem $\mathcal{S}_i$ along any of the $d$ branches, we can follow the branch and thereby encounter infinitely many $\mathcal{S}_j$. Because there are only finitely many $\mathcal{S}_j$ in $\mathcal{S}$, we encounter one $\mathcal{S}_j$ at least two times, i.e. the path is closed in the forward direction (with respect to the coverings). The same argument applies to the backward direction, so that every path can be completed to one single closed path within $\mathcal{S}$ along any branch.


\section{Constrained subsystems}
\label{sec:constr}

\begin{figure}[hbtp]
\begin{center}
\includegraphics{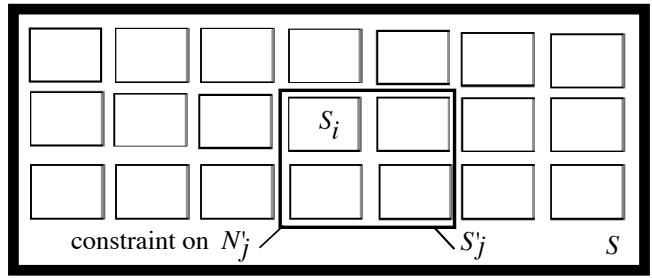}
\end{center}
\caption[Partitions under constraints]{\label{fig:5}Partitions under constraints}
\end{figure}

Up to this point, we have considered equilibrium partitions and uniform distributions. Uniform distributions are not very exciting. However, we can obtain additional information on the numbers $N'_j$ for certain subsystems $\mathcal{S}'_j$ of $\mathcal{S}$, due to observational data or due to symmetry conditions imposed on the parameterisation (local invariances under translations or rotations). The additional informations are constraints which must be imposed while maximising the entropy of $\mathcal{S}$.
As a consequence, $\mathcal{S}$ no longer has a uniform distribution. The constraints are typically imposed at subscales of $\mathcal{S}$ which are larger than $\mathcal{S}_i$. The overall situation is summarised in figure \ref{fig:5}.

\paragraph*{}
Observations can be performed on the numbers $N'_j$ of quanta, but not on the ordering, for the following reason. The ordering is an arbitrary mathematical construction whereas the primary quanta are given independently of any mathematical construction. Therefore, any ''measurable physical phenomenon'' may depend on the counting of the quanta, but not on how the construction of an ordering is performed. Thus, no observational data constraints can be imposed on $\{m_i\}$. Because the parameterisation symmetry constraints do not depend on $i$, $\{m_i\}$ cannot be constrained at all. It follows that $\mathcal{P}$ is equilibrium with respect to the dimension, and thus $m_i = d =$ constant. By the same argument as in section \ref{sec:param}, every path can thus be closed. 

In order to evaluate the distribution $(\{N_i\}, \{n_i\})$ of $\mathcal{S}$, we have to maximise its entropy $S$ while imposing a certain number $M$ of constraints $g_q(\mathcal{P}) = c_q = {\rm constant}$, $q=1\ldots M$, and $g_q$ refers to a system $\mathcal{S}'_j(q)$.This yields: 

\begin{equation}
\label{eq:SecondLaw}
\big[\delta S + \delta \sum_{q=1}^M \lambda_q \ (g_q - c_q)\big]_\mathcal{S} = 0
\end{equation}

\noindent with Lagrange multipliers $\lambda_q$. Equation (\ref{eq:SecondLaw}) can be written as a sum over $\mathcal{J}=\{i|\mathcal{S}_i \subset \mathcal{S}\}$:

\begin{eqnarray}
\sum_{i\in \mathcal{J}}\delta S_i + \delta \sum_{q=1}^{M} \lambda_q \ (g_q - c_q) & = & \nonumber \\
\label{eq:S_Sum}
\sum_{i\in \mathcal{J}}\delta [s(x_i^k)\prod_{k=1}^d \Delta x_i^k] + \delta \sum_{q=1}^{M} \lambda_q \ (g_q - c_q) & = & 0,
\end{eqnarray}

\noindent where $s(x_i^k)$ is the $d$-density of $S$ and is defined by (\ref{eq:S_Sum}). For large $\mathcal{J}$, (\ref{eq:S_Sum}) can be converted to an integral expression with the domain $\mathcal{T}  = x(\mathcal{S})$:

\begin{equation}
\label{eq:S_Int}
\delta S_c = \delta \oint_\mathcal{T} {\rm d}^d x \ [s(x^k) + \sum_{l=1}^{m_c} \ \lambda_l(x^k) \ \zeta_l(x^k)] = 0
\end{equation}

\noindent with $m_c$ Lagrange multiplier ''functions'' $\lambda_l(x^k) = \lambda_q$ and constraint functions $\zeta_l(x^k)$ defined by 

\begin{eqnarray}
\lambda_{q} \ (g_q - c_q) & = & \int_{x(\mathcal{S}'_j(q))} {\rm d}^d x \ \lambda_l(x^k) \ \zeta_l(x^k), \nonumber \\
\label{eq:constraints}
\sum_{q=1}^{M} \lambda_{q} \ (g_q - c_q) & = & \sum_{l=1}^{m_c} \oint_\mathcal{T} {\rm d}^d x \ \lambda_l(x^k) \ \zeta_l(x^k).
\end{eqnarray} 

\noindent The closed integral in (\ref{eq:S_Int},\ref{eq:constraints}) arises because all paths can be closed, and thus the integral may also be interpreted as a boundary integral. Therefore, the range of integration must have either a periodicity or a periodic identification of end points. On the other hand, $x^k$ is arbitrary, i.e. $x^k$ is not required to satisfy any periodicity conditions at all. This is unsatisfactory if $x^k$ is taken to be real. Conversely, we can substitute ${\rm d}^dx = {\rm  i}{\rm d}^dx_L$ in the integral expression to obtain $S_c = {\rm  i}\oint {\rm d}^d x_L [\ldots] = {\rm  i}S_{cL}$ and, therefore, $\Omega = \exp [{\rm  i}S_{cL}]$ exhibits a periodic identification of points modulo $2\pi$. This transformation is nothing else than a ''Wick rotation''. If we can find a foliation of hypersurfaces $x_L^d =$ constant so that $x^d={\rm  i}x_L^d$, then $x_L^d$ is analogous to the time component of space-time. $x^d=x^d_E$ has Euclidean, $x^k_L$ has Lorentzian-like ''signature''.

\paragraph*{}
The analytical properties of the $x^k$-dependent functions depend on the type of parameterisation we apply. The observational constraints could imply singular points. In the present article, however, we shall only consider well-behaved parameterisations and constraints, i.e. the integrands $s(x^k)$ and $\lambda_l(x^k) \ \zeta_l(x^k)$ are smooth functions of $x^k$ after maximisation of $S_c$ ($\delta S_c = 0$). Therefore, we can apply the tools of differential geometry.

\paragraph*{}
Because $\mathcal{T}$ may be interpreted as a boundary, it is possible to apply Gauss' law on (\ref{eq:S_Int}). To this end, it is useful to substitute the freely variable parameterisation $x^k$ by one fixed choice $\bar{x}^k$. We interpret  ${\rm d}^d\bar{x}$ as the volume form: it is not affected by a change of the parameterisation $x^k$. By contrast, the factors ${\rm d}^d x$ and $s$ change whenever we perform a coordinate transformation $x^k \rightarrow x'^k$. In order for $s$ to remain smooth after this transformation, $\partial \bar{x}^j / \partial x^k$ itself must be smooth. It even must be a diffeomorphism in order to be invertible. We must require the transformations $x^k \rightarrow x'^k$ to be invertible so that we do not loose information under the back-transformation. The entropy does not change under a transformation and is thus diffeomorphism invariant. Let us use the simplest fixed parameterisation, which is the cartesian, $(\bar{x}^K) \in R^d$, $K = 1 \ldots d$. Introducing the vielbeins $e^K_k = \partial \bar{x}^K / \partial x^k$ and the (Euclidean) metric $\gamma_{kl} = e^K_k \eta_{KL} e^L_l$ with determinant $\gamma$ yields 

\begin{eqnarray}
& \delta \oint_\mathcal{T} {\rm d}^d \bar{x} &  [\gamma^{-1/2} \ s \ + \ \gamma^{-1/2} \sum_{l=1}^{m_c} \ \lambda_l \ \zeta_l] = \nonumber \\
\label{eq:S_vielbein}
& \delta \oint_\mathcal{T} {\rm d}^d x & \sqrt{\gamma} \ [\gamma^{-1/2} \ s + \sum_{l=1}^{m_c} \ \lambda_l \ \xi_l] = 0,
\end{eqnarray}

\noindent where $\xi_l(x^k) = \gamma^{-1/2}\zeta_l(x^k)$.
Due to the parameterisation, the variation also includes the vielbeins and functions thereof. We do not only distribute the quanta randomly among the boxes, but also with respect to $x^k$. Because the parameterisation is arbitrary, $\sigma = \gamma^{-1/2} s$ transforms as $\sigma'(x'^k) = \sigma(x^k)$ under translations $x'^k = x^k+a^k$ and does not explicitly depend on $x^k$.
Define

\begin{equation}
\label{eq:tau_kappa}
\tau^k_K = \gamma^{-1/2} \ s \ \gamma^{kl}e^L_l \ \eta_{LK}, \quad
\kappa = \sum_{l=1}^{m_c} \ \lambda_l \ \xi_l.
\end{equation}

\noindent (\ref{eq:S_vielbein}) can then be expressed shortly as

\begin{equation}
\label{eq:S_tau}
\delta \oint_\mathcal{T} {\rm d}^d x \ \sqrt{\gamma} \ [\tau^k_K \ e^K_k + \kappa] = 0.
\end{equation}

\noindent If $\mathcal{T}$ is orientable, (\ref{eq:S_tau}) can be converted into an integral over a ($d+1$)-dimensional volume $\mathcal{M}$ with boundary $\partial \mathcal{M} = \mathcal{T}$. To construct $\mathcal{M}$, each boundary element is supplemented by an (outward) unit vector $(n^\gamma) = (0,\ldots,0,1)$ normal to the local space $V(x^k)$. The boundary element can then be written as ${\rm d}\sigma_\gamma = n_\gamma {\rm d}^d x$. The integrand of (\ref{eq:S_tau}) must be converted into a ($d+1$)-dimensional vector with extended notation $\tau^k_K \ e^K_k = \tau^{k\gamma}_K \ e^K_k \ n_\gamma$ and $\kappa = \kappa^\gamma \ n_\gamma$ (up to a new gauge freedom due to the index $\gamma$). We then extend the embedding of $\mathcal{T}$ to the bulk $\mathcal{M}$ as a smooth manifold with extended cartesian fixed parameters $\bar{x}^\mu$ and not yet determined vielbeins $e^\Delta_\delta$, metric $g_{\mu\nu}$ (determinant $g$) and possibly torsion. With this trick, we can apply Gauss' theorem:

\begin{eqnarray}
\oint_\mathcal{T} {\rm d}^d x \ n_\gamma \sqrt{\gamma} \ [\tau^{k\gamma}_K \ e^K_k + \kappa^\gamma] & = &  \nonumber \\
\label{eq:gauss}
\int_\mathcal{M} {\rm d}^{d+1} x \ \sqrt{g} \ [\nabla_\gamma (\tau^{\delta\gamma}_\Delta \ e^\Delta_\delta) + \nabla_\gamma \kappa^\gamma],
\end{eqnarray}

\noindent If $\mathcal{T}$ fails to be orientable, there are closed paths $\Gamma: t \in [0,1] \rightarrow x_t^k \in \mathcal{T}$ along which the initial (positive) orientation is flipped, $\det[d\{x_{t\rightarrow 1}^j(x_0^k)\}]=-1$. A special coordinate basis is $x_t^1 = t$, $x_1^2=-x_0^2$, $x_1^{k>2}=x_0^k$. Around $\Gamma$, $\mathcal{T}$ looks like a moebus-strip of dimension $d$, it has non-vanishing torsion. However, $S = \sum_i S_i$ has the same value as for the untwisted space $\tilde{\mathcal{T}}$ obtained from $\mathcal{T}$ by setting $x_1^2= +x_0^2$. The presence of constraints reduces the number of microstates and thus the ''net entropy'', $S \rightarrow S_c$. Thus, $\tilde{S}_c(\tilde{\mathcal{T}}) = S_c(\mathcal{T})$ and $S_c$ is also left unchanged. Because $\tilde{\mathcal{T}}$ is orientable, Gauss' theorem applies. Thus, we only need to replace $\nabla_\gamma$ by the torsionless covariant derivative $\tilde{\nabla}_\gamma$ in (\ref{eq:gauss}). Let us finally perform the Wick-rotation $x^\mu \rightarrow x^\mu_L$ and drop the index $L$ in what follows. The first term of (\ref{eq:gauss}) can be expanded:

\begin{eqnarray}
& \rho & = \tilde{\nabla}_\gamma(\tau^{\delta\gamma}_\Delta \ e^\Delta_\delta ) = \delta_\mu^\delta \tilde{\nabla}_\gamma(\tau^{\mu\gamma}_\Lambda \ e^\Lambda_\delta ) \nonumber \\
& = & g_{\mu\nu} g^{\nu\delta} \tilde{\nabla}_\gamma(\tau^{\mu\gamma}_\Lambda \ e^\Lambda_\delta ) \nonumber \\
& = & e^\Delta_\mu \eta_{\Delta\Gamma} e^\Gamma_\nu g^{\nu\delta} \tilde{\nabla}_\gamma (\tau^{\mu\gamma}_\Lambda \ e^\Lambda_\delta ) \nonumber \\
\label{eq:expand_rho}
& = & e^\Delta_\mu e^\Gamma_\nu \Phi^{\mu\nu}_{\Delta\Gamma},
\end{eqnarray}

\noindent where 

\begin{equation}
\label{eq:Phi}
\Phi^{\mu\nu}_{\Delta\Gamma} = \eta_{\Delta\Gamma} g^{\nu\delta} \tilde{\nabla}_\gamma [e^\Lambda_\delta \tau^{\mu\gamma}_\Lambda]. 
\end{equation}

\noindent Equation (\ref{eq:S_tau}) then becomes

\begin{equation}
\label{eq:motion}
\delta \int_\mathcal{M} {\rm d}^{d+1} x \ \sqrt{-g} \ [e^\Delta_\mu e^\Gamma_\nu \Phi^{\mu\nu}_{\Delta\Gamma} + \mu] = 0.
\end{equation}

\noindent with $\mu = \tilde{\nabla}_\mu \kappa^\mu$. (\ref{eq:expand_rho}) has a similar form as the Palatini action. $\Phi^{\mu\nu}_{\Delta\Gamma}$ plays a similar role as the curvature two-form. Depending on the structure of $\mu$, there may be some torsion contribution, via the connection one-form

\begin{equation}
\label{eq:omega}
\omega_{\mu \Delta\Gamma}  = e^\alpha_\Delta \nabla_\mu e_{\alpha \Gamma}
\end{equation}

\noindent if it has a component symmetric in the capital indices.

\paragraph*{}
In the same way as $\sigma$ does not depend explicitly on $x^k$, $\mu$ does not depend explicitly on $x^\mu$. 
$\mu$ is the sum of the contributions from the parameterisation symmetry and the observational data constraints, $\mu = \mu_{\rm psc} + \mu_{\rm odc}$.


\section{Torsion-less case and GR}
\label{sec:torsionless}

\paragraph*{}
Consider (\ref{eq:expand_rho}) in the special case of vanishing torsion (when sources of torsion are absent or their effects cancel each other at small scale). Choosing a cartesian fixed parameterisation $\bar{x}^K$ (see section \ref{sec:constr}) yields the metricity, $\nabla_\gamma g_{\alpha\beta} = 0$. With $\tau^{\alpha\beta\gamma} = \tau^{\alpha\gamma}_\Gamma \ \eta^{\Gamma\Delta} \ e_\Delta^\beta$, we thus have

\begin{equation}
\label{eq:leibnitz2}
\rho = \nabla_\gamma(\tau^{\alpha\beta\gamma} \ g_{\alpha\beta}) 
= \rho^{\alpha\beta} \ g_{\alpha\beta},
\end{equation}

\noindent where Leibnitz' rule and

\begin{equation}
\label{eq:Sigma}
\rho^{\alpha\beta} = \tilde{\nabla}_\gamma\tau^{\alpha\beta\gamma} = \nabla_\gamma\tau^{\alpha\beta\gamma}
\end{equation}

\noindent have been used. $\rho^{\alpha\beta}$ explicitly depends on $g_{\mu\nu}$ and derivatives thereof. It plays a role analogous to the Ricci tensor $R^{\alpha\beta}$ in GR. Let us therefore define the tensor which is analogous to the Einstein tensor $G_{\mu\nu}$:

\begin{equation}
\label{eq:chi}
\chi_{\mu\nu} = \rho_{\mu\nu} - \rho g_{\mu\nu} / 2.
\end{equation}

\noindent On the other hand, in order to obtain tensorial field equations, let us define the variational derivative

\begin{equation}
\label{eq:theta}
\theta^{\mu\nu} = \frac{2}{\sqrt{-g}} \ \frac{\delta (\sqrt{-g} \mu)}{\delta g_{\mu\nu}}.
\end{equation}

\noindent Because $g_{\mu\nu}$ is symmetric in its indices, we are lead to

\begin{eqnarray}
& & \delta \int_\mathcal{M} {\rm d}^{d+1} x \ \sqrt{-g} \ [\rho + \mu] \nonumber \\
& = & \int_\mathcal{M} {\rm d}^{d+1} x \ [\rho_{\mu\nu} \ \delta (\sqrt{-g} \ g^{\mu\nu}) + \sqrt{-g} \ \theta_{\mu\nu} \delta g^{\mu\nu}] \nonumber \\
& + & \int_\mathcal{M} {\rm d}^{d+1} x \ \sqrt{-g} \ g^{\mu\nu} \ \delta \rho_{\mu\nu} \nonumber \\
& = & \int_\mathcal{M} {\rm d}^{d+1} x \ \sqrt{-g} \ [\chi_{\mu\nu} + \theta_{\mu\nu}] \ \delta g^{\mu\nu} \nonumber \\
\label{eq:motion_torsionless}
& + & \int_\mathcal{M} {\rm d}^{d+1} x \ \sqrt{-g} \ g^{\mu\nu} \ \delta \rho_{\mu\nu} = 0.
\end{eqnarray}

\noindent The variation of the second but last line can be performed after having fixed the last line (divergence form) to zero and inserted the resulting quantities into the second but last line. We obtain the field equations

\begin{equation}
\label{eq:field_eq}
\chi^{\mu\nu} = - \theta^{\mu\nu}.
\end{equation}

\paragraph*{}
It shall now be shown that $\chi_{\mu\nu}$ is divergence-free. We can express the variation $\delta g^{\mu\nu}$ from a transformation $x^\mu \rightarrow x^\mu + \epsilon a^\mu(x^\nu)$ with infinitesimal $\epsilon$,

\begin{equation}
\label{eq:delta_gmunu}
\delta g^{\mu\nu} = \epsilon \ (g^{\mu\lambda}\nabla_\lambda a^\nu + g^{\nu\lambda}\nabla_\lambda a^\mu),
\end{equation}

\noindent and let $\theta^\lambda_\nu a^\nu$ vanish on the boundary (divergence form contribution) to obtain the following expression via Euclidean coordinates:
 
\begin{eqnarray}
0 & = & \int_\mathcal{M} {\rm d}^{d+1} x_E \ \frac{\delta (\sqrt{g_E} \mu)}{\delta g^{\mu\nu}} \ \delta g^{\mu\nu} \nonumber \\
& = & \int_\mathcal{M} {\rm d}^{d+1} x_E \ \sqrt{g_E} \ \theta_{\mu\nu} \ \delta g^{\mu\nu} \nonumber \\
& = & \epsilon \int_\mathcal{M} {\rm d}^{d+1} x_E \ \sqrt{g_E} \ \theta^\lambda_\nu \ \nabla_\lambda a^\nu \nonumber \\
\label{eq:thetamunu_conserv}
& = & -\epsilon \int_\mathcal{M} {\rm d}^{d+1} x_E \ \sqrt{g_E} \ a^\nu \ \nabla_\lambda \theta^\lambda_\nu.
\end{eqnarray}

\noindent Because $a^\nu$ is arbitrary and after turning back to Lorentz coordinates $x_L$, it follows that $\theta^{\mu\nu}$ is divergence-free and, from (\ref{eq:field_eq}), $\chi^{\mu\nu}$ is divergence-free as claimed.

\paragraph*{}
Because the form of (\ref{eq:field_eq}) is similar to the form of Einstein's equation, let us compare both within the currently observable range of gravitational curvature. This range of curvature is sufficiently well described if we consider contributions no higher than quadratic in the dimension of derivatives of $g_{\mu\nu}$. In the case $d=3$, it follows from \cite{Lovelock} that the only possible form of $\chi_{\mu\nu}$ is

\begin{equation}
\label{eq:Einstein_approx}
\chi_{\mu\nu} = A G_{\mu\nu} + B g_{\mu\nu},
\end{equation}

\noindent where $A$ and $B$ are constants, $G_{\mu\nu}$ is computed  starting from $g_{\mu\nu}$ using the formula for the Einstein tensor and $B$ leaves the possibility for a cosmological constant $\Lambda$.
With (\ref{eq:Einstein_approx}), GR has been shown to be dual to the non-dynamical approach for $d=3$, $A$ and $B$ appropriately given by GR, without torsion and up to the observationally relevant second order in the dimension of the derivatives of $g_{\mu\nu}$. However, no explanation for the condition $d=3$ and the constant values $A$ and $B$ are given here.

\paragraph*{}
Because gravitational space and the non-dynamical parameter space are dual to each other, we can extend the duality to the matter: the quantity $\mu$ is the analogue of the matter Lagrangian $\mathcal{L}_m$. Let us therefore introduce matter field functions $\psi_l(x^k)$ which play a similar role for matter dynamics as the ''gravitational'' field $e^\Gamma_\gamma$ for gravitational dynamics. $\psi_l$ should therefore be written in a quadratic form, as formal functions with appropriate definition of the squared norm $|\psi_l|^2 = \psi_l^\dagger \psi_l = C_l$ and conveniently chosen normalisation $C_l$.
With $\pi_l^\gamma$ defined by $\pi_l^\gamma \psi_l n_\gamma = \lambda_l \ \xi_l$, we obtain

\begin{eqnarray}
\label{eq:psi}
\kappa^\gamma n_\gamma & = & \sum_l \lambda_l \xi_l = \sum_l \pi_l^\gamma \psi_l n_\gamma, \\
\label{eq:mu_psi}
\mu & = & \sum_l [(\nabla_\gamma \pi_l^\gamma) \ \psi_l + \pi_l^\gamma \ \nabla_\gamma \psi_l].
\end{eqnarray}


\section{Quantum measurements and QFT}
\label{sec:quantum_theory}

\begin{figure}[hbtp]
\begin{center}
\includegraphics{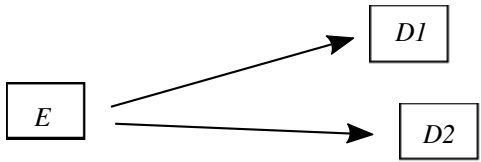}
\end{center}
\caption[Quantum emitter and detectors]{\label{fig:6}Quantum emitter and detectors}
\end{figure}

Let us first sketch a typical quantum process, where quantum mechanical information (e.g. a particle) is emitted by an emitter $E$ and detected by a detector $D1$ or $D2$ $\ldots$, see figure \ref{fig:6} for two detectors. For simplicity, we assume that one detector fires. One of the $n_D$ detector systems $\mathcal{S}_{D1}, \mathcal{S}_{D2}, \ldots, \mathcal{S}_{Dn_D}$ will thus undergo a change of its quantum number distribution $(\{N_i\},\{n_i\})$ in such a way that it relaxes to a new state of maximum entropy. 

The question one can ask is: What is the probability that the quantum information is detected by $D1$ rather than any other $Dk$ ($k \ne 1$)? The answer is given by the constrained numbers of possible states, $\Omega(E|Dk)$ for ''emitted in $E$ and detected in $Dk$'':

\begin{equation}
\label{eq:prob} 
p(E\rightarrow D1) = \frac{\Omega(E|D1)}{\sum_{k=1}^{n_D}\Omega(E|Dk)},
\end{equation}

\begin{eqnarray}
& & \Omega(E|Dk) = {\rm e}^{(iS_c\big|_{E|Dk})} = \nonumber \\
\label{eq:OmegaDk} 
& & \exp[i(S\big|_{E|Dk} + \int_\mathcal{M} {\rm d}^{d+1} x \ \sqrt{-g} \ \mu_{E|Dk})].
\end{eqnarray}

\noindent In order to evaluate (\ref{eq:prob}), we need to solve \mbox{$\delta S_c\big|_{E|Dk} = 0$}. This is a potentially very large set of coupled inhomogeneous nonlinear differential equations involving $S[e^\Delta_\mu, \omega_{\mu\Delta\Gamma}]$ and $\sqrt{-g}$. Specific strategies for solving this equation will be analysed elsewhere.

\paragraph*{}
Let us finally address the question: Can QFT in flat space-time be recovered? In the flat ($d+1$)-parameter space approximation (with cartesian flat coordinates), we set $e^\Gamma_\gamma \approx \delta^\Gamma_\gamma$, $g_{\mu\nu} \approx \eta_{\mu\nu}$, $\omega_{\mu\Delta\Gamma} \approx 0$ and hence $\exp({\rm i} S) \approx 1$. Naively, one would be tempted to write

\begin{equation}
\label{eq:wrong}
\Omega \approx e^{{\rm i} S_m} \quad \rightarrow \delta S_m \approx 0,
\end{equation}

\noindent where $S_m = \int_\mathcal{M} {\rm d}^{d+1} x \ \mu$. However, the flat parameter space is an infinite domain of integration and does not admit a boundary, so that the expression $S_m = \int_{\partial \mathcal{M}} {\rm d}^d x_E \ \sqrt{\gamma_E} \ \kappa_E$ is ill-defined. Therefore, (\ref{eq:wrong}) cannot be used. We must consider a finite volume $V$ and treat it as an open system. Let us assume that no observational data constraint is imposed, $\mu_{\rm odc} = 0$. Then, only the parameterisation symmetry constraint remains, $\mu = \mu_{\rm psc}$, which depends on $e^\Gamma_\gamma$ and $\omega_{\mu\Delta\Gamma}$.
Let us also choose $V$ small enough so that all quantities can be approximated by constants as a function of the location within $V$. This is possible because all quantities are smooth functions of $x^\mu$. Then, $V$ behaves as a system in thermal equilibrium and exchanging heat with a thermal bath (its neighbourhood). Therefore, we must replace the (microcanonical) number of states $\Omega$ by the (canonical) partition function

\begin{eqnarray}
& & Z(V, T) \approx \int [\prod_{\Gamma,\gamma} {\rm d} e^\Gamma_\gamma] \ [\prod_{\mu,\Delta,\Lambda} {\rm d} \omega_{\mu\Delta\Lambda}] \ {\rm e}^{{\rm i} [E(V)/T+S_m(V)]} \nonumber \\
\label{eq:Z_e}
& & = \int [\prod_{\Gamma,\gamma} {\rm d} e^\Gamma_\gamma] \ [\prod_{\mu,\Delta,\Lambda} {\rm d} \omega_{\mu\Delta\Lambda}] \ {\rm e}^{{\rm i} S_c(V,T)}.
\end{eqnarray}

\paragraph*{}
In the approximation of a weak field $|e^\Gamma_\gamma - \delta^\Gamma_\gamma| \ll 1$ and weak torsion, i.e. $|\omega_{\mu(\Delta\Lambda)}| \ll 1$ with the notation $X_{\mu(\Delta\Lambda)} = (X_{\mu\Delta\Lambda}+X_{\mu\Lambda\Delta})/2$, we can expand $e^\Gamma_\gamma =$ \mbox{$\delta^\Gamma_\gamma + \epsilon a^\Gamma_\gamma(x^\mu)$}, $\omega_{\mu(\Delta\Lambda)} = \omega\epsilon b_{\mu(\Delta\Lambda)}(x^\mu)$, assuming $\epsilon \ll 1$. 

\paragraph*{}
In the case $\mu=0$, the saddle-point $\delta S = 0$ is given by $\epsilon = 0$. This means $\partial S / \partial \epsilon\big|_{\epsilon=0} = 0 = \partial \rho / \partial \epsilon\big|_{\epsilon=0}$ and therefore $\rho(\epsilon) = \mathcal{O}(\epsilon^2)$ in the vicinity of $\epsilon=0$. However, because of (\ref{eq:expand_rho}), $\rho(\epsilon)$ must possess a non-vanishing term $\sim \epsilon^2$, i.e. $\rho \sim  \epsilon^2 + \mathcal{O}(\epsilon^3)$.

\paragraph*{}
In the expression (\ref{eq:Z_e}), only values ${\rm e}^{{\rm i} S_c}$ near the saddle-point contribute significantly to $Z(V,T)$. Thus, for $\mu \ne 0$, we can consider $S_c$ near the saddle-point, approximate $\mu_{\rm psc} \sim \epsilon^2 + \mathcal{O}(\epsilon^3)$ and vary $\epsilon$ while holding $\rho$ fixed. Let us assume that $\mu_{\rm psc}$ only contributes with terms $\sim |\psi_l|^2$ to $e^\Gamma_\gamma$ and $\omega_{\mu\Delta\Gamma}$, while interaction terms (e.g. $\sim g\psi_k \psi_{l\ne k} \ldots$) are small perturbations. One single field $\psi$ yields $\psi \sim \epsilon$. For several fields $\psi_l$, the contributions of the $\psi_l$ to $\epsilon$ decouple from each other, and we can write $\psi_l \sim \epsilon_l$, so that $e^\Gamma_\gamma \approx \delta^\Gamma_\gamma + \sum_l \epsilon_l (a_l)^\Gamma_\gamma$ and $\omega_{\mu(\Delta\Lambda)} \approx \sum_l \omega_l\epsilon_l (b_l)_{\mu(\Delta\Lambda)}$. This yields:

\begin{eqnarray}
& & Z(V, T) \approx \int [\prod_{\Gamma,\gamma} {\rm d} e^\Gamma_\gamma] \ [\prod_{\mu,\Delta,\Lambda} {\rm d} \omega_{\mu(\Delta\Lambda)}] \nonumber \\
& & \cdot \ {\rm e}^{{\rm i} \int_V {\rm d}^{d+1}x \ \sqrt{-g}\ (\rho + \mu_{\rm psc})} \approx 
c \int [\prod_{l} {\rm d} \epsilon_l] \ [\prod_{l} {\rm d} (\omega_l\epsilon_l)] \nonumber \\
\label{eq:Z_psi}
& & \cdot \ {\rm e}^{{\rm i} \int_V {\rm d}^{d+1}x \ \mu_{\rm psc}} \approx c' \int [\prod_{l} {\rm d} \psi_l] \ {\rm e}^{{\rm i} \int_V {\rm d}^{d+1}x \ \mu_{\rm psc}},
\end{eqnarray}

\noindent where, in the second but last expression, $\sqrt{-g} \approx 1$, $\rho \approx 0$, $\mu_{\rm psc}$ does not depend on the orientation of $(a_l)^\Gamma_\gamma$ and $(b_l)_{\mu(\Delta\Lambda)}$, $c$ and $c'$ are constants and we finally only have to keep ${\rm d} \epsilon_l = {\rm d} \psi_l$ in the measure of each gravitational contribution of $\psi_l$, while the torsion contribution $\omega_{\mu(\Delta\Lambda)} \sim \omega_l\epsilon_l \ll 1$ has been neglected in $\mu_{\rm psc}$ and the measure ${\rm d} (\omega_l\epsilon_l)$ drops out.
After summing over all small volumes $V_k$ within a finite but large volume $V_{tot} = L^4$, we end up with the total partition function $Z(V_{\rm tot})$ which has the form of the Feynman path integral representation of QFT for $d=3$ and $L\rightarrow \infty$:

\begin{eqnarray}
& & Z(V_{tot}) \approx \lim_{k\rightarrow \infty} \prod_k Z(V_k,T_k) \nonumber \\
\label{eq:interm}
& & = \int [\prod_l \mathcal{D}\psi_l] \ \exp({\rm i}\int_{V_{\rm tot}} {\rm d}^{d+1}x \ \mu_{\rm psc}).
\end{eqnarray}

\noindent The NDA model thus allows to recover a quantum behaviour fully analogous to QFT for $d=3$. $\hbar \mu_{\rm psc}$ then corresponds to the Lagrangian of the matter fields $\psi_l$. We can identify $\hbar$ as one unit of entropy, i.e. one box if we fix $p-1=1$. On the other hand, a large number of quanta per volume corresponds to a high temperature. It may be conjectured that a high temperature is related to a high density of matter and, from (\ref{eq:motion}), that the amount of $E$ per primary quantum is related to the Planck energy. This conjecture will be examined elsewhere.

This section has only given a brief insight to how the recovery of quantum theory can be obtained.


\section{\label{sec:conclusion}Conclusions}

Starting from a large set of objects and no quantum dynamics, a general mathematical derivation (statistics of systems of objects) has allowed to obtain a model in a form analogous to GR in case of 3-dimensional parameterisation (up to experimental limits of observation) and analogous to the Feynman path integral approach to QFT in the flat space approximation (assuming the interaction terms to be small). Due to the absence of a Lagrangian or Hamiltonian at quantum level, no splitting between space and time occurs for DNA, in contrast to the canonical quantisation method and in contrast to curved QFT. Further effort is still required in order to obtain explicit solutions. Further work should also clarify the reason for the dimension $d=3$, the structure of matter fields and interactions (complete systematics even beyond the standard model by construction) and further development towards the explicit formulation of a quantum model of gravity.



\begin{acknowledgments}
I would like to thank Gino Isidori for fruitful discussions and Philippe Jetzer for hospitality at University of Zurich.
\end{acknowledgments}

\bibliography{ndaconcept}

\end{document}